\RequirePackage{fix-cm}

\documentclass[smallcondensed]{svjour3}

\usepackage{graphicx}
\usepackage{subfigure}
\usepackage{natbib}
\usepackage[center]{caption}

\journalname{Celestial Mechanics and Dynamical Astronomy}

\begin{document}
%%%%%%%%%%%%%%%%%%%% HEADER
   \title{A study of low-energy transfer orbits to the Moon: towards an operational optimization technique}

	\author{R. Capuzzo-Dolcetta \and M. Giancotti}
	\institute{Roberto Capuzzo-Dolcetta
               \at Dep. of Physics, Sapienza, Univ. of Rome,
               \\\email{Roberto.Capuzzodolcetta@uniroma1.it}
               \and
               Marco Giancotti
               \at Scuola di Ingegneria Aerospaziale, Sapienza Univ. of Rome,
               \\\email{marco.giancotti@uniroma1.it}}
	\date{Received: date / Accepted: date} % The correct dates will be entered by the editor
	\maketitle

%%%%%%%%%%%%%%%%%%%% ABSTRACT
	\begin{abstract}
      In the Earth-Moon system, low-energy orbits are transfer trajectories
      from the earth to a circumlunar orbit that require less propellant consumption
      when compared to the traditional methods. In this work we use a Monte
      Carlo approach to study a great number of such transfer orbits over a
      wide range of initial conditions. We make statistical and operational
      considerations on the resulting data, leading to the  description of a
      reliable way of finding ``optimal'' mission orbits with the tools of
      multi-objective optimization.  \keywords{Low-energy Orbits \and N-Body
      \and Restricted Problems \and Spacecraft \and Numerical Methods}
	\PACS{95.10.Ce \and 95.55.Pe}
	\end{abstract}

	%%%%%%%%%%%%%%%%%%%% SECTION
	\section{Introduction}
	\label{intro}
   When facing the problem of transferring a spacecraft from an orbit around
   the Earth to one around the Moon, the option usually taken is that of
   a two-impulse  Hohmann transfer. The past two decades have seen the
   discovery and development of a new family of orbits that exploit the $N$-body
   dynamics to reduce the $\Delta v$ budget of interplanetary missions.
   These ``low-energy orbits'' are rather new but have already been used
   several times to reach the Moon and the Earth-Sun libration points. Examples
   are the Japanese Hiten, the European SMART-1 and the NASA GRAIL missions.

   The purpose of this work is to study how the different types of transfer orbits
   to the Moon compare to each other in terms of performance and room occupied
   in the space of the free parameters like the elements of the initial orbit and
   the launch date. This is done with a Monte Carlo approach, producing a random
   population of transfer orbits within a wide range of initial conditions.

   Section \ref{theory} briefly describes the dynamics of the $3$- and $4$-body
   problem which make the low-energy orbits possible. These are divided into
   the ``outer'' type and the ``inner'' type, rather different from each other.
   Section \ref{method} describes the way the numerical simulation was performed,
   the initial conditions and the algorithm used.
   Section \ref{results} contains an analysis of the $\Delta v$, the total duration
   and their dependence on the initial conditions of the transfer orbits.
   Section \ref{opti} describes how the data could be analysed with the methods
   of multi-objective optimization to search for the preferable type of
   transfer for a mission.
   Section \ref{concl} states the conclusions.

	%%%%%%%%%%%%%%%%%%%% SECTION
   \section{Types of Earth-Moon Transfers}
   \label{theory}
   \subsection{The N-Body Problem}
      The study of the dynamics of a small body in the Solar System corresponds to
      searching the solution of a classical, Newtonian, $N+1$-body problem, where $N$
      is the number of gravitating bodies (Sun and planets) in the system, with given
      initial conditions.
      
      The problem of flying a probe in the Solar
      System is that of a point of negligible mass that moves along the geodesic
      defined by its initial conditions in the time varying potential given by the
      revolving planets and the Sun. A precise integration of the equations of motion
      of the probe, even in the simplifying assumption of point mass particle, is
      obtainable by numerical methods once the right hand side (forces) is given
      through high precision ephemerides. The transfer of a spacecraft from
      the Earth to the Moon can be studied, at a high level of accuracy, as that of a
      body of negligible mass in the field of 3 finite mass bodies (Earth, Moon and
      Sun).  Analytical solutions are only available in the case of
      triangular and rectilinear configurations. For generic applications, numerical
      methods are unavoidable.
      
      The requirements characterizing the PCR3BP are:
      \begin{itemize}
         \item one of the three bodies is much smaller than the other two
               and its gravitational effect on them is negligible;
         \item the two larger bodies, which we will call the \emph{primaries}
               $A$ and $B$ travel on circular orbits around each other;
         \item the third body, which we will call the \emph{particle} $P$, moves
               in the plane of the primaries.
      \end{itemize}
      In particular, Joseph Louis Lagrange's
      work on the $3$-body problem led him to the discovery of five points of
      equilibrium, sometimes called \emph{libration
      points}, on which the whole system's dynamics is based.

      The energy and momentum of the third body ($P$)
      are not conserved quantities
      when $N=3$. The only first integral known in the PCR3BP is the
      \emph{Jacobi integral} $C$, defined as
      \begin{equation}
         C = -2\Phi(x,y) - (\dot{x}^2 + \dot{y}^2)\,,
      \end{equation}
      where $\dot{x}$ and $\dot{y}$ are the components of the velocity of $P$
      and $\Phi(x,y)$ is the effective potential, containing both gravitational and
      centrifugal terms.
      For any given value of $C$, the motion of the particle is spatially confined
      inside the so called \emph{Hill region},
      \begin{equation} \label{eq:restr3_hill}
         \left\{ (x,y)\in R^{2}:\:C\leq-2\Phi(x,y) \right\}\,,
      \end{equation}
      bounded by \emph{zero velocity curves} (ZVC), which have different
      shapes depending on the different values of $C$ and of the system's masses.

      \begin{figure*}
         \centering
         \subfigure[]{
            \label{subf:grad_good}
   	      \includegraphics[width=0.45\textwidth]{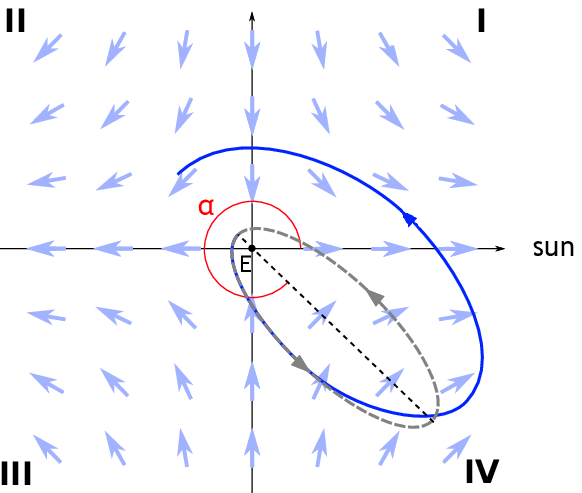}
         }
         \subfigure[]{
            \label{subf:grad_bad}
   	      \includegraphics[width=0.45\textwidth]{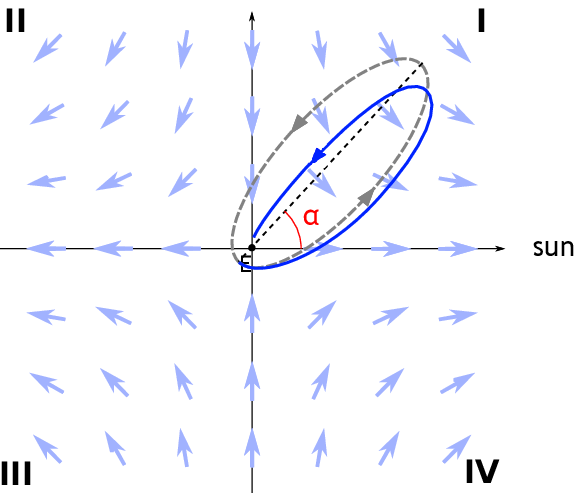}
         }
   	\caption{The $\alpha$ angle and the shape of the gravity gradient induced by the Sun
      in the surroundings of the Earth. The direction of the Sun and the numbers
      of the quadrants (I, II, III, IV)
      are shown. \subref{subf:grad_good} Deviation from the (initially osculating) Kepler
      orbit (dashed gray line) when $\alpha$ is in
      an even quadrant: the final perigee is higher than the initial one.
      \subref{subf:grad_bad} When $\alpha$ is in an odd quadrant, the length of the major axis
      is reduced, leading to a lower perigee.} 
   	\label{fig:grad}
   	\end{figure*}
      \subsection{Lunar Transfer and Low Energy Orbits}
      In the specific case of a transfer to the Moon, the two primary bodies in
      the $3$-body formulation of the problem are the Earth and the Moon,
      while the spacecraft ($P$) is the third body. This system does not
      meet the requirements of the PCR3BP exactly. However it is close enough
      to display qualitatively equivalent solutions.

      The traditional method of bringing a payload from the Earth to the Moon is through a Hohmann
      transfer which has a typical total duration $T_t$ of $5$ days (from LEO).
      This procedure requires a value of $C$
      lower than that of the $4^{th}$ and $5^{th}$ libration points, thus selecting
      a configuration of the Hill region where motion is allowed
      in the whole space, without forbidden regions. In this case the duration of the
      flight and the energy of the system allow us to ignore the complex dynamics tied
      to the libration points. Apart from later corrections for the various perturbations,
      this approach only makes use of $2$-body celestial mechanics principles.

      Recent research has led to the discovery of another family of trajectories
      exploiting the properties of the Hill region and of the libration points
      \citep{belbruno-internal, belbruno-hiten,kawaguchi1995,kolomaro2000, kolomaro2001}.
      In order to make this possible, $C$ needs to be close to the critical values, confining
      the space of possible motion
      and thus allowing the spacecraft to be ``captured'' in the
      area around the Moon. Some of these trajectories can be achieved only if the Jacobi
      ``constant'' $C$ varies
      throughout the flight. The intervention of a fourth body is therefore needed
      to provide an external perturbation to the $3$-body system. These are sometimes
      called ``low-energy orbits''.
   
      \paragraph{Influence of the Sun.} The gradient of the solar gravitational
      field in the surroundings of the Earth is expressed, neglecting the second
      order terms, with the tidal tensor
      \begin{equation}
         \mathbf{\Phi} = \frac{GM_{sun}}{R^3}\left[3\hat{R}\hat{R} - \mathbf{u}\right]\mathbf{r},
      \end{equation}
      where $\mathbf{R}$ is the Sun-Earth distance vector, $\mathbf{u}$ the unit
      dyadic and $\mathbf{r}$ is the Earth-spacecraft distance vector. The field
      introduces a force pushing $P$ away from the Earth along
      the Sun-Earth line and towards it in the direction orthogonal to it, as shown
      in Figure \ref{fig:grad}. The key parameter relating this effect to the dynamics
      of an elliptical orbit is the angle, $\alpha$, between the Sun-Earth line
      and the orbit's major axis \citep{teofilatto2001}. When $\alpha$ is inside the
      $2^{nd}$ or $4^{th}$ quadrants in the figure,
      the Sun's gravity field aids the orbit by rising the spacecraft's energy,
      especially when it is close to apogee (Figure \ref{fig:grad}a.). The opposite
      happens when $\alpha$ is in one of the other
      two quadrants, where the spacecraft's energy is lowered (Figure \ref{fig:grad}b.).
      The overall effect is an increase or decrease of the orbit's major axis growing
      non-linearly with its length.

      It was found \citep{belbruno-hiten} that it is possible to exploit this phenomenon
      in order to design spacecraft trajectories that reach the Moon with lower
      relative kinetic energy, thus decreasing the overall fuel requirements
      of the mission. The spacecraft is ``captured'' gravitationally by the lunar
      field in a process called a \emph{ballistic capture}.

      \paragraph{Ballistic capture.} Traditional Hohmann-type transfers approach
      the Moon with a relatively high velocity requiring the use of on-board rockets
      to slow down the motion of $P$ until it becomes bounded to the Moon. In order
      to obtain a stable, bound orbit around the Moon, the energy of the spacecraft
      with respect to it must be decreased to negative values.  In the case of
      the $3$-body problem, the condition is $C>C_1$, where $C$ is
      the Jacobi constant of the spacecraft and $C_1$ is that related to the first libration point $L_1$.
      This corresponds to closing the $L_1$ neck of the Hill region so that passage
      between the Earth and Moon domains of possible motion is blocked and $P$ is confined
      in the region around the Moon.

      With ballistic capture, the spacecraft approaches the Moon with a
      low relative velocity and enters directly into a temporarily bound orbit.
      An example of ballistic capture in nature is that of
      some comets in the combined Sun-Jupiter gravitational field.
      Most famous is comet 39P/Oterma, which was captured for a short time around
      Jupiter in 1936.

      It must be noted that the chaotic $3$-body dynamics of the system allows, generally, 
      for a temporary (unstable) ballistic capture, although it is thought that permanent capture 
      (i.e. the probe remains bound to the Moon forever) can happen \citep{belbruno-capture}.
      Normally, after the ballistic capture
      transfer, the spacecraft performs one or more revolutions around the Moon
      before escaping again into one of the other realms of possible motion. For this
      reason an active maneuver is necessary to stabilize into a lunar orbit. The clear
      advantage of using this type of
      trajectory lies in its lower fuel requirements with respect to usual Hohmann transfers.

      \paragraph{Outer transfer.} The contribution of the Sun described above can
      be used to naturally modify the shape of a highly eccentric orbit. This is the
      case already shown in Figure \ref{fig:grad}. When the
      apogee distance $r_a$ is close to $1.5\times10^6\textnormal{km}$, the effect
      becomes strong enough to raise the successive perigee up to about
      $4\times10^5\textnormal{ km}$. When this happens, the spacecraft can
      approach the Moon with a lower relative velocity than a Hohmann transfer,
      possibly resulting in a ballistic capture. In this paper this type of
      trajectory is called \emph{outer transfer}.

      The advantage of this type of transfers, when ballistic capture occurs,
      is the reduced $\Delta v$ in the final phase, when propulsion is needed to stabilize the orbit. The data
      analyzed in the following paragraphs shows that this method results in
      a reduction of up to $44\%$ of this second impulse compared to that of a
      Hohmann transfer.

      \paragraph{Inner resonance transfer.}
      A second type of low-energy transfers does not make use of the effect
      of the Sun. Instead, it is based on the unstable dynamics near the
      Earth-Moon $L_1$ libration point. In this case, which we will call
      \emph{inner resonance transfer} or inner transfer, the apogee distance $r_a$
      produced by the initial $\Delta v$ lies inside the orbit of the Moon, roughly
      $3.30\times10^5\textnormal{ km}$ from the Earth. The duration of the transfer $T_t$ goes from
      $85$ days upwards, depending on how many Earth orbits are completed
      before falling into the Moon's area of influence.

      In these transfers the
      effect of the Sun is much weaker and it doesn't help
      in \emph{raising the apogee} up to the lunar distance, even
      during the ``favorable'' $\alpha$ quadrant epochs. Yet its contribution can
      be important when $\alpha$ is in one of the other two quadrants,
      because it can \emph{lower the perigee} enough to make the spacecraft
      impact the Earth.

      The process by which the capture occurs despite the ``short apogee'' is
      the execution of one or more \emph{resonance hops} with
      respect to the Moon. Each time the spacecraft flies by $L_1$ its trajectory
      can shift into a new near-resonant state, usually with a longer major
      axis. This is repeated until the spacecraft is able to enter the Moon's sphere
      of influence. The dynamics behind these resonance hops is tied
      to the structure of phase space when $P$ is close in energy and position
      to $L_1$ \citep{belbruno1997,belbruno2008}. However its
      details are still not completely understood.

      \paragraph{Interpretation}
      The explanation given above involving tidal forces and the $\alpha$ angle
      \citep{teofilatto2001} is only one of several. The original rigorous interpretation
      of the phenomenon is given in \citet{belbruno-internal}, \citet{belbruno1993},
      \citet{bello-mora2000}
      through the concept of weak stability boundary. This is the region
      where the gravitational forces due to the various bodies are similar, allowing
      for ballistic capture and propellant savings.

      Another insightful interpretation involves the stable and unstable manifolds
      related to the periodic orbits at the collinear libration points
      \citep{Conley1968,kolomaro2000,kolomaro2001}. The phase space position of
      the spacecraft with respect to these two manifolds determines its qualitative
      behavior in the framework of the $3$-body problem.
      By choosing carefully the initial conditions of the orbit it is possible
      to build complex paths going around several celestial objects with few
      or no active maneuvers.

	%%%%%%%%%%%%%%%%%%%% SECTION
	\section{Method}
	\label{method}
   	The approach followed in this work is statistical, through the search of 
      generic transfer orbits from a low Earth parking orbit to a stable lunar orbit.
      The nature of the
   	simulation makes it possible to study the complete phenomenology of the transfer
   	problem, eliminating any technical or theoretical bias that could be included
   	in an ad-hoc algorithm. The aim is to contribute a slightly more complete picture
      of the phenomenology of the transfers, rather than discovering new methods to
      create low-energy orbits.
   
   	The simulations are performed using a restricted version of the Solar System consisting
   	of Earth, Moon and Sun. They are modeled, in the basic settings, as point masses with
      realistic, non-planar motion, taken from the ephemerides
      of the Jet Propulsion Laboratory's Horizons System\footnote{http://ssd.jpl.nasa.gov/?horizons}. 
      The other planets are not considered in the present research because they
      introduce negligible effects in the short term.

\subsection{Initial Conditions for the Orbits}
      Each transfer is made of two separate ignitions:
      $\Delta v_1$ to begin the transfer at time $t_0$ and $\Delta v_2$, at time $t_2$, to 
      inject the spacecraft into a stable orbit around the Moon.
      The initial condition \emph{for all the transfers} is a circular low Earth
      parking orbit with a radius $r_0=6720\textnormal{ km}$, %(Table \ref{tab:orb})
      and $\Delta v_1$ is always applied in a direction parallel to the instantaneous velocity
      \footnote{Note that a high trust propulsion capable of instantaneous bursts is assumed.}.

      The ecliptic coordinate system $(x_{ec},y_{ec},z_{ec})$ is assumed for the
      rest of this paper, unless differently stated.
      The inclination of the plane of the (circular) parking orbit can be varied using two angular
      parameters: $\theta$, the angle of rotation around the $x_{ec}$ axis
      and $\nu$, the angle of rotation around the $y_{ec}$ axis. The angle of rotation
      around the $z_{ec}$ axis is never changed. 

      Being the parking orbit circular, the argument of perigee is not defined,
      and the initial conditions for the satellite motion leaving the parking
      orbit are given on the crossing of the parking orbit plane and the
      ecliptic. This is the only strong restriction given to the initial conditions.
      It greatly reduces the parameter space that is explored, but also makes
      the number of necessary calculations manageable.

   	$\Delta v_1$ is chosen randomly in an interval between a specified minimum
      value (different choices were tested) and a maximum value that corresponds to the 
   	escape from the Earth's gravitational well starting from the parking orbit, 
      i.e. $\Delta v_1=3.19\textnormal{ km s}^{-1}$.
      After the boost, the trajectory is
   	integrated forward in time for a period of 160 days. Among the orbits thus
   	produced, all those that do not enter the sphere of radius $10^4$ km around
      the Moon are discarded. Only those passing closer to the Moon than $10^4$ km
      without colliding with it are stored.
   
   	After this first screening the algorithm applies a 2-impulse Hohmann maneuver
      to reach a stable $2000$ km circular orbit around the Moon (an altitude of $262$ km).
      The first impulse of this maneuver is performed at the point closest to the Moon
      of the $160$-day orbit. The portion of the trajectory
      that comes \emph{after} this point of closest approach is discarded. The result
      is an elliptic orbit with periselenium distance of $2\times10^3$ km. The
      other impulse is done at periselenium and it circularizes the orbit. In total
      these two maneuvers require a variation of velocity indicated as $\Delta v_2$.

      \begin{figure*}
      \centering
      \subfigure[]{
         \label{subfig:energy}
         \includegraphics[width=0.45\textwidth]{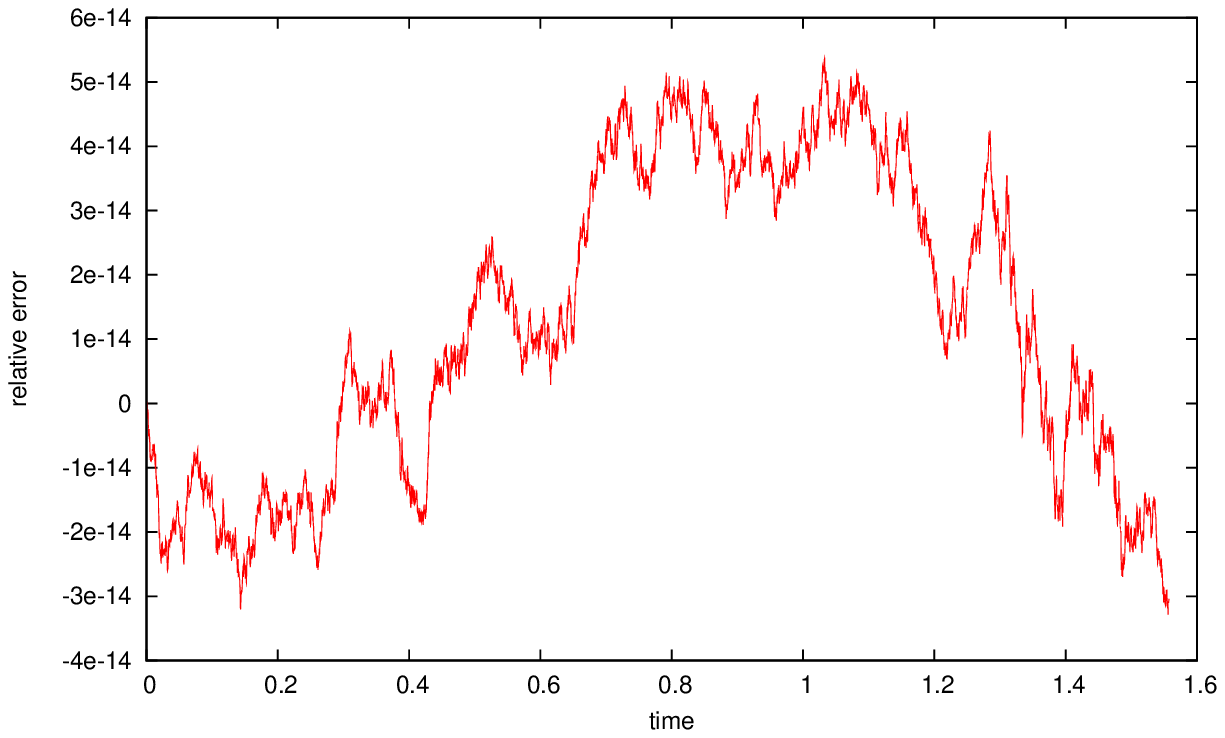}
            }
      \subfigure[]{
         \label{subfig:errorstep}
         \includegraphics[width=0.45\textwidth]{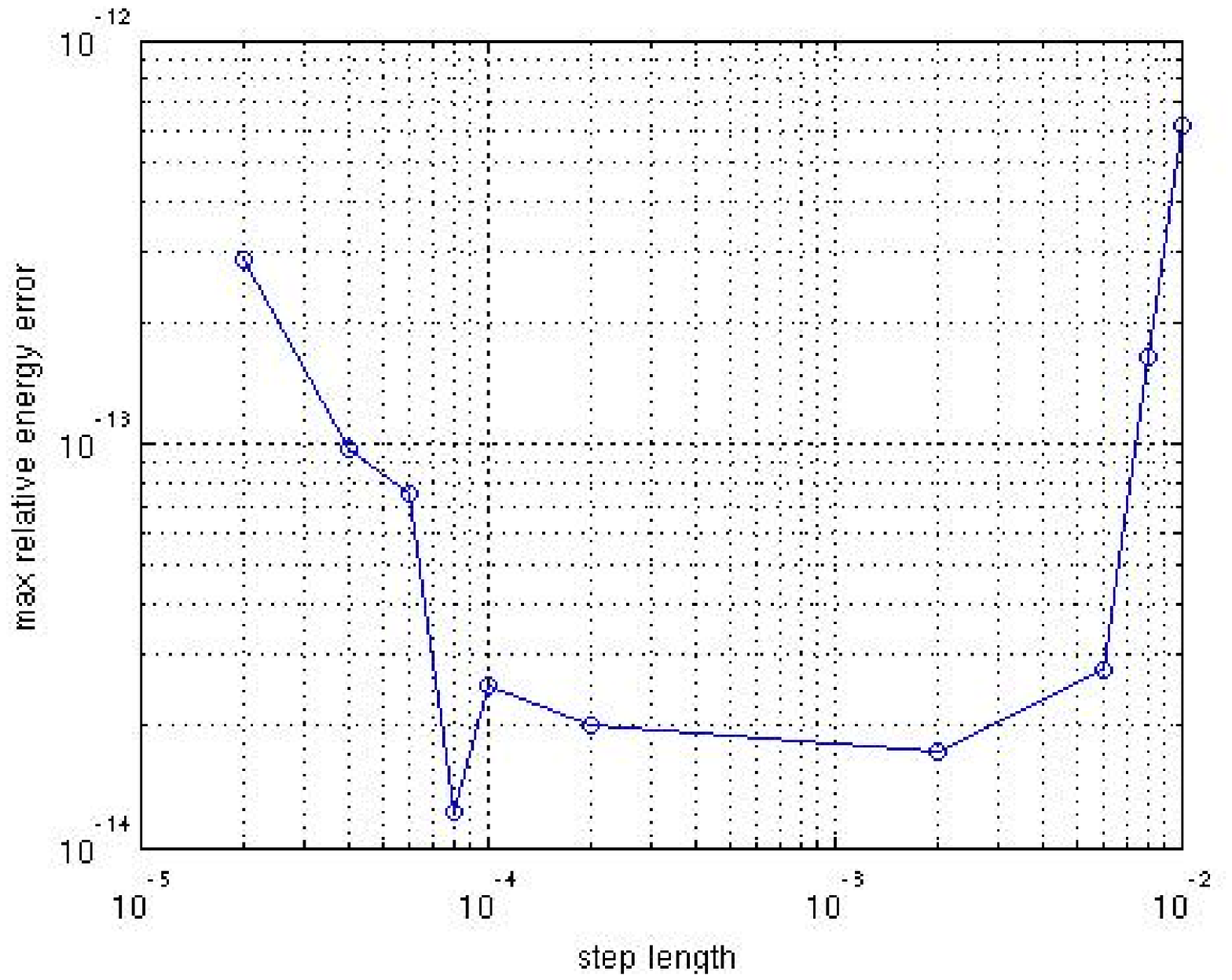}
            }
            \caption{\subref{subfig:energy}: relative error in energy in the
            orbital integration for an orbit lasting $93$ days (time is in
            system's units). \subref{subfig:errorstep}: maximum relative error
            in energy over a full orbit, relative to different lengths of the
            time step.} \label{fig:energy_ab}
      \end{figure*}
\subsection{Orbital Integration}
      The integration of the equations of motion of a spacecraft submitted to a
      variable external potential with given initial conditions must be of the
      highest possible precision. Anyway, it is well known that ordinary numerical
      methods for integrating Newtonian equations of motions become dissipative and
      exhibit incorrect long term behavior (see, e.g., \citet{men84,mka90,car92}).
      This is a serious problem performing $N$-body computations, particularly when
      studying the evolution of systems over large time intervals.

      One possible solution is using symplectic integrators, which are a particular
      type of {\it geometric} integrators which, respect to ordinary integrators,
      have the advantage to be qualitatively better because they preserve the
      physical properties of the original, although discretized, system's
      hamiltonian. This allows for a good conservation of the characteristics of the
      dynamical system's time flow properties, thing that is particularly appreciated
      when the simulation is performed over an extended time, with an
      energy error much better constrained than with ordinary methods which show an
      irreversible drift. The advantage of using a symplectic algorithm of
      sufficiently high order is that it gives more reliability of the quality of the
      results at a computer time consumption similar to that of a non-symplectic
      method, so to make it worth using also for integrations limited in time.

      The choice of symplectic methods is wide, for it is possible to construct high
      order integrators \citep{yos90}. For the purposes of this paper, we used the
      $6^{th}$-order explicit scheme whose coefficients are taken from the first
      column of the Table 1 (SI6A) of  \citet{Kin91}, which leads to a conservation
      of energy by a factor $50$ better than with the other two possible sets of
      coefficients reported in the same Table. The integration of the spacecraft
      orbits in the external field was done by mean of a properly adapted version of
      the full N-body code NBSymple, developed by \citet{cap11}. The code was also
      adapted to include a multipolar expansion (up to the 4th degree)
      for the Sun and the Earth potential and to account for the influence of
      the whole set of planets in the solar system, but these features were
      not used for the final computations.

      Although the potential of the Sun is conveniently represented by a truncated
      multipolar expansion, accounting for its deviation from spherical symmetry, it
      is easily seen that the contribution of the quadrupole and hexadecapole (P$_2$
      and P$_4$ Legendre's terms, respectively) contribute for a fraction of only
      about $7\times 10^{-12}$ to the solar gravitational force at one Astronomical
      Unit. Also the lunar potential, at the
      distances from the Moon of interest for this paper, is sufficiently well
      represented by the monopole term, as can be seen by comparing it to higher moments
      as deduced from the Lunar Prospector mission by \citet{han11}.  So in the
      following calculation we considered, as absolutely sufficient for the scopes of
      this paper, just the solar, Earth and Moon monopole term in their gravitational
      potential.

      The orbit of the point-like spacecraft is integrated with a
      maximum relative error of $2\times 10^{-14}$ (see Figure \ref{fig:energy_ab}). 
      When the time step is within the range $10^{-4} \leq \Delta t \leq 3\times
      10^{-3}$, our code, working on a standard
      Intel Xeon CPU is able to generate in one minute $\sim 250$ orbits with a
      duration of 160 days.

      Further details on the numerical N-body code structure and characteristics can
      be found in \citet{cap11}.

	%%%%%%%%%%%%%%%%%%%% SECTION
\section{Results}
\label{results}
      The method employed in the simulation explores a wide portion of the
      parameter space of the Earth-Moon transfer problem. Several kinds of
      orbits linking the two astronomical bodies are found. As it might be
      expected, most of the trajectories found are 
      of limited applicability, with very large $\Delta v$ and long durations.
      However, a significant number of ``effective''
      trajectories, i.e. viable to practical utilization, exists. Besides
      the traditional Hohmann-like transfers, they also include the
      two types of low-energy transfers that exploit the $3$- and $4$-body
      dynamics of the Earth-Moon-Sun system (inner and outer types, described above).
   
      A total of $5733$ complete transfer trajectories were produced with the
      following $4$ main simulation modes:
      \begin{enumerate}
      \item Progression of dates: several transfers each day during the course of one
         year. Lower limit on $\Delta v_1$ is such that the minimum $r_a$ of the transfer
         orbit is $r_{a,min}=3.3\times10^5\textnormal{ km}$. This includes the Hohmann
         transfers and the inner and outer low-energy transfers. Angles $\nu$ and $\theta$
         are both $0$.
      \item Progression of dates: like point 1, but the lower limit on $\Delta v_1$ is such that
         $r_{a,min}=8\times10^5\textnormal{ km}$. This \emph{excludes} the Hohmann and inner
         transfers.
      \item progression of values for $\nu$; lower limit on $\Delta v_1$ such that
         $r_{a,min}=8\times10^5\textnormal{ km}$.
      \item progression of values for $\theta$; lower limit on $\Delta v_1$ such that
         $r_{a,min}=8\times10^5\textnormal{ km}$.
      \end{enumerate}
      The choice of parameters is aimed both at exploring a wide spectrum
      of possibilities and at increasing the number of low-energy orbits to
      study.
      Most of the orbits were produced with the first two modes.

      \subsection{Discussion on the Relation Between $\Delta v_1$ and $\Delta v_2$}
      	\begin{figure*}
            \centering
        \includegraphics[width=1.0\textwidth]{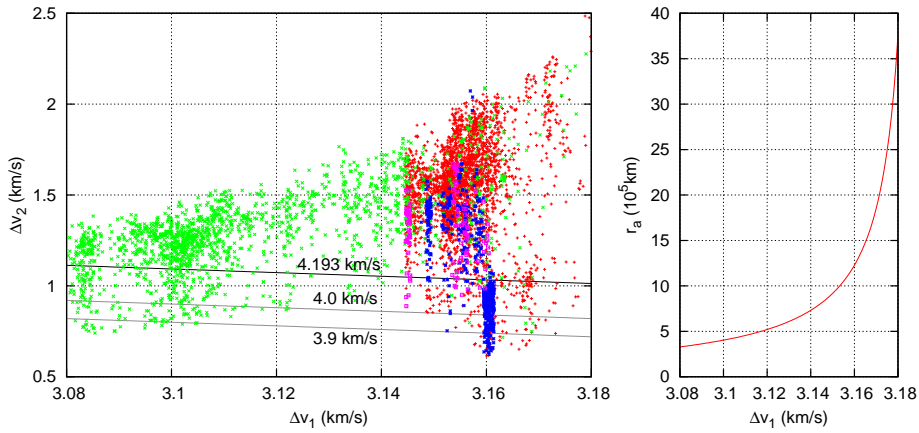}
      	\caption{Left: distribution on the $\Delta v_1$-$\Delta v_2$ plane of $5733$
            transfers. Green markers: simulation mode A (see text); red: mode B;
            magenta: mode C; blue: mode D. Lines corresponding to the equation
            $\Delta v_{tot} = \Delta v_1+\Delta v_2 = k$ for three different
            values of $k$ (as labeled) are shown.
            Right: relation between $r_a$ of an elliptical Kepler orbit
            and $\Delta v_1$, given the initial conditions employed in the simulation.}
      	\label{fig:dv12}
      	\end{figure*}
      
         A first aspect of the data produced by the simulations is
         the dependence of the second impulse ($\Delta v_2$) on
         the first ($\Delta v_1$). $\Delta v_1$ is only allowed to vary between
         $3.08\textnormal{ km s}^{-1}$ and  $3.19\textnormal{ km s}^{-1}$ (the value
         necessary to reach the escape velocity), which results in apogee distances
         ranging from $r_{a,min}=3.3\times10^5\textnormal{ km }$ to
         infinity (the extreme case being a parabolic orbit).
         $\Delta v_2$, on the other hand, ranges roughly from $0.7$ to $2.0\textnormal{ km s}^{-1}$.
         
         The left panel of Figure \ref{fig:dv12} shows a collection of orbits
         produced for different simulation modes. 
         On this plot, an average Hohmann trajectory is located in the proximity
         of the point $[3.100\textnormal{ km s}^{-1}$, $1.093\textnormal{ km s}^{-1}]$.
         Different markers denote the orbits produced with different modes (see figure
         caption). Note that the magenta and blue points are not as spread out as the
         other two modes because
         they were all simulated starting from the same initial date, during a favorable
         quadrant of $\alpha$.

         The curve in the right panel of Figure \ref{fig:dv12} shows, for reference,
         the relation between $r_a$ and $\Delta v_1$ given the initial conditions in use.
         Note that the Earth-Moon distance is $\sim 3.85\times10^5$ km and
         the apogee of an outer orbit is $r_a=1.2\sim\!1.5\times10^6$ km.
         
         \paragraph{Main strip.}
         The distribution in Figure \ref{fig:dv12} displays a direct positive
         correlation between $\Delta v_1$ and $\Delta v_2$ up to $\Delta v_1=3.140\textnormal{ km s}^{-1}$, which
         corresponds to $r_a=7.4\times10^5\textnormal{ km}$. In this region the
         distribution takes the shape of a strip with increasing values of $\Delta v_2$.
         
         For larger values of $\Delta v_1$, 
         the distribution becomes widely scattered, branching also towards lower
         values of $\Delta v_2$. Most of the transfers on the main strip, excluding
         those in the lower-right area, are inefficient versions of Hohmann transfers and many
         of them perform several revolutions before coming close to the Moon.
         The reason for this ``Hohmann strip'' in Figure \ref{fig:dv12} is straightforward,
         best explained by separating the cases in which $\Delta v_1$ is
         greater than or less than that of the theoretical Hohmann which we will
         here call $\Delta v_1^h$:
         \begin{enumerate}
            \item $\Delta v_1>\Delta v_1^h$: here the transfer orbits cross
            the lunar orbit and go beyond it for a while instead of being
            tangent to it. Trajectories resulting from higher $\Delta v_1$'s
            will therefore approach the
            Moon with wider angles with respect to the lunar instantaneous velocity.
            The higher this approach angle is, the higher $\Delta v_2$ needs to be
            in order to attain a stable final orbit;
            \item $\Delta v_1<\Delta v_1^h$: a traditional Hohmann transfer is
            not possible because the elliptic orbit does not intersect
            or touch the Moon's trajectory.
            However the passage of $P$ close to the first libration point activates
            complex $3$-body dynamics that can lead to a lunar orbit. In particular,
            most of these orbits perform resonance hops, increasing the length of
            their major axis and eventually falling into a ballistic capture around
            the Moon. The resulting $\Delta v_2$ used to stabilize the orbit is usually lower than
            that of a Hohmann transfer, and in some cases similar to
            that of an outer transfer.
         \end{enumerate}
         
         \paragraph{Distribution for higher $\Delta v_1$.} For $\Delta
         v_1>3.140\textnormal{ km s}^{-1}$, the line of the distribution in
         Figure \ref{fig:dv12} expands also to low values of $\Delta v_2$. The
         upper portion is a continuation of the previously described Hohmann strip,
         while the lower branch is composed of low-energy outer transfers.

         The difference between the two branches can be entirely ascribed to
         the $4$-body dynamics due to the Sun, which becomes important
         when the point of apogee reaches the implied large distances from the Earth.
         The upper branch is dominant when the $\alpha$ angle between the Sun, the
         Earth and the orbit's apogee
         is unfavorable, i.e. in the $1^{st}$ or $3^{rd}$ quadrants.
         Conversely, when $\alpha$ is in the $2^{nd}$ or $4^{th}$ quadrants, the
         probability of obtaining an outer low-energy transfer is higher, producing
         the lower branch.

         All the points of the lower branch around and below the $\Delta v_{tot}=4.0\textnormal{ km s}^{-1}$
         diagonal line represent outer low-energy transfers. The lowest value reached is
         $\Delta v_2\simeq0.61\textnormal{ km s}^{-1}$, that is, $44\%$ less than
         the $\Delta v_2$ of the \emph{average} Hohmann transfer.

      \subsection{Transfer Duration and $\Delta v$}

      	\begin{figure*}
            \centering
            \includegraphics[width=0.9\textwidth]{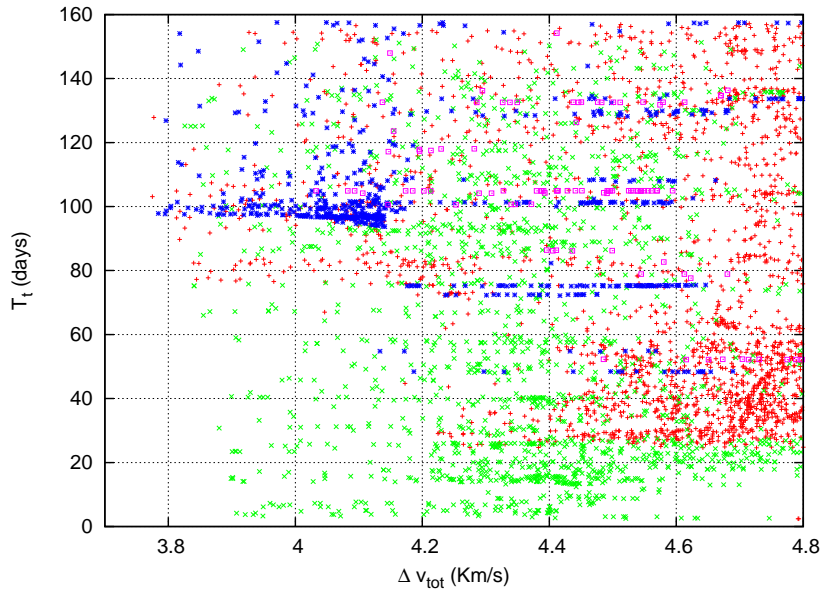}
            \caption{$5733$ transfers in the $\Delta v_{tot}$-$T_t$ plane.
               The same symbols as Figure \ref{fig:dv12} apply.}
      	   \label{fig:comptot}
      	\end{figure*}

         Two of the most relevant parameters to be taken into account
         in the choice of the transfer orbit for a
         mission are the total propellant consumption, expressed through $\Delta v_{tot}$,
         and the total time of flight, $T_t$. Depending on the specific purpose of the
         mission, greater importance may be given to one parameter over the other.
         The propellant used grows exponentially with the $\Delta v$
         of a maneuver, and from the point of view of the budget it should
         be kept as low as possible. On the other hand, missions that take a long time to
         reach the target can be unacceptable for various reasons, like life
         support costs in manned missions and related risks. 

         The plot in Figure \ref{fig:comptot} shows the distribution of the orbits
         in the $\Delta v_{tot}$-$T_t$ plane. The orbits are produced with different
         choices of the $t_0$, $\nu$ angle and of the lower limit to $r_a$ (displayed
         in the plot with different markers). Figure \ref{fig:comptot} can be used
         as an aid in the choice of an optimal trajectory given a specific mission, a topic
         explored in more detail in Section \ref{opti}.
         
      \subsection{Dependence on the Launch Date} \label{dv2alpha}
         The data obtained from the simulation show that the date of launch influences
         the $\Delta v$ of the transfer orbits. The Moon
         needs to be in an appropriate part of its orbit in order to be reached by the
         spacecraft, and the required position is different for the various types of transfers.
         The dependence on the lunar cycle is evident in the obtained data from the oscillation
         of the probability of transfer with a period of $27.3$ days, i.e. the period
         of the Moon's revolution around the Earth.

      	\begin{figure*}
            \centering
            \includegraphics[width=0.9\textwidth]{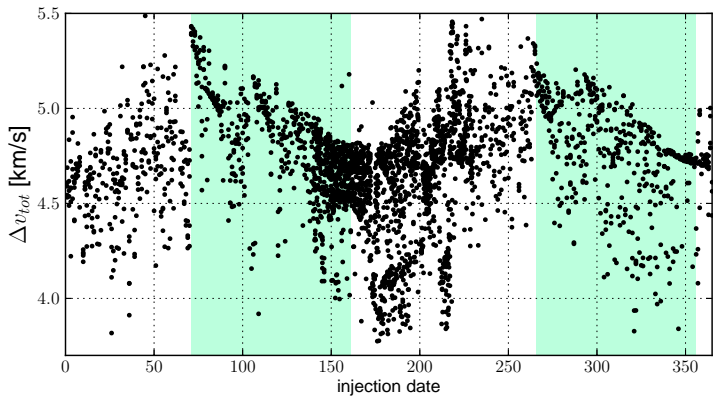}
            \caption{Dependence on the launch date for a set of orbits with
            $\Delta v_1>3.14\textnormal{ km s}^{-1}$. The ``unfavorable'' quadrants
            ($1^{\textnormal{st}}$ and $3^{\textnormal{rd}}$) are shown on a
            darker background.}
            \label{fig:alphadep}
      	\end{figure*}
         The second major effect is given by the angle $\alpha$ between the apogee
         and the Sun (described
         in Sect. \ref{theory}). The gradient of the Sun's gravitational field
         around the Earth's position
         results in a tidal force acting on the orbiting body, whose direction depends on the
         angular configuration of the Sun-Earth-spacecraft system.

         For short-term orbits, like the traditional Hohmann transfers taking $3$ to
         $5$ days to reach the Moon, the consequences of this force are negligible.
         However the contribution of the Sun cannot be overlooked when considering
         trajectories with $r_a$ greater than that of the Moon, with $T_t\ge80$ days.
         By changing the shape of the orbits, the Sun affects the velocity with
         which the spacecraft passes close to the Moon. The result is
         a variation in the $\Delta v_2$ needed to enter a stable lunar orbit.

         This is visible in Figure~\ref{fig:alphadep}, where $\Delta v_{tot}$ for
         different launch dates is shown along with the periods in which $\alpha$ is
         in its various quadrants. Note that the date shown for the points is when
         $\Delta v_1$ is applied; a spacecraft in a low-energy orbit comes
         close to apogee, feeling the effect of the Sun, after roughly $30$ to $70$
         days. Therefore, despite the appearance, the crests and troughs of the
         distribution do correspond to the predicted $\alpha$ phases \citep{teofilatto2001}.
         Also, the most efficient orbits of the distribution are all located in the favorable
         quadrants of $\alpha$, as will be shown in Section \ref{opti}, Figure
         \ref{fig:pareto_date}. 

         Of all the transfers with $\Delta v_2<1.0\textnormal{ km s}^{-1}$, $93\%$ have the
         apogee angle $\alpha$ inside either
         the $2^{nd}$ or the $4^{th}$ quadrant. The
         transfers happening when $\alpha$ is at the boundary between two quadrants
         ($\alpha=k\,\pi/2$ with $k=1,2,3,4$) display
         deformed shapes, because of the different effects of the Sun's
         gradient along the orbit.

         Inner resonance transfers, with $r_a<4\times10^5\textnormal{ km}$ do not
        show a clear dependence on $\alpha$. The mechanism leading
         to a ballistic capture is, in this case, different and mostly independent 
         of the position of the Sun.
         
      \subsection{Low-Energy Orbits}
      	\begin{figure*}[b]
            \centering
            \includegraphics[width=0.7\textwidth]{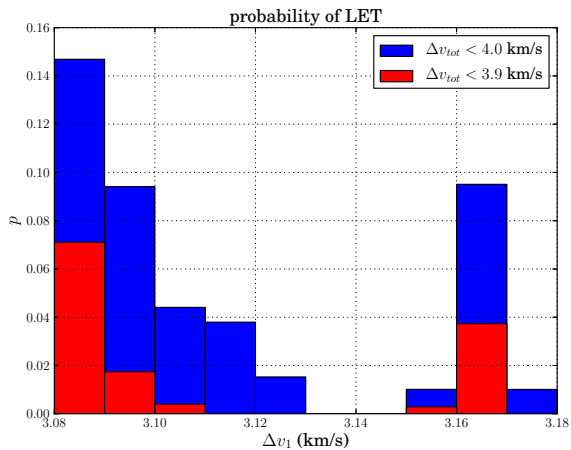}
            \caption{Histogram showing the portion of orbits that are
               low-energy, using two alternative definitions based on
               $\Delta v_{tot}$.}
      	   \label{fig:hist-prob}
      	\end{figure*}
         \paragraph{Existence of low-energy orbits.}
         Taking into account different initial conditions (i.e. the
         A, B, C and D modes described earlier)
         the number of orbits found to be low-energy transfers constitutes a
         non negligible fraction of the total.
         The probability of producing a low-energy transfer when shooting
         in a given interval of $\Delta v_1$ is represented in the histogram
         in Figure \ref{fig:hist-prob}. There is no rigorous definition of
         low-energy transfer, but putting an upper limit on the $\Delta v_{tot}$
         allows us to exclude the Hohmann and the other high-cost transfers.
         Considering that the average Hohmann transfer in this study has
         $\Delta v_{tot} = 4.193\textnormal{ km s}^{-1}$, we can improvise
         two alternative definitions, one requiring $\Delta v_{tot} <
         4.0\textnormal{ km s}^{-1}$ and another, stricter one requiring
         $\Delta v_{tot} < 3.9\textnormal{ km s}^{-1}$.

         Figure \ref{fig:hist-prob} shows the portion of all the orbits, in each
         bin of $\Delta v_1$, that fall into these two definitions. In both
         cases the low energy orbits exist in two separate groups, as expected
         from Section \ref{theory}: one for low values (inner transfers) and
         one for high values (outer transfers) of $\Delta v_1$. For both definitions
         there exists a gap between the two groups where no conforming transfer
         takes place.
  
         The distribution of the transfers in $\Delta v_1$ can be translated into ranges
         of apogee distances. The inner transfers thus have apogee distances approximately
         $3.36\times10^5\textnormal{ km} < r_a < 3.43\times10^5\textnormal{ km}$.
         Outer transfers, on the other hand, exist in the range
         $1.23\times10^6\textnormal{ km} < r_a < 1.31\times10^6\textnormal{ km}$.
         Most of the optimized orbits described in other works go further out,
         up to $1.5\times10^6\textnormal{ km}$ from the Earth. The fact that these
         longer orbits are rare in this (non-optimized) simulation
         is certainly due to the greater difficulty in producing them. They occupy a
         smaller volume in the space of the parameters and therefore our Monte Carlo
         approach is less likely to find them. Of course, this more sensitive dependence
         on the initial conditions is the trade-off necessary to obtain higher
         efficiencies.

      	\begin{figure*}
            \centering
            \subfigure[]{
               \label{subf:lowen-hists-in-dv}
      	      \includegraphics[width=0.45\textwidth]{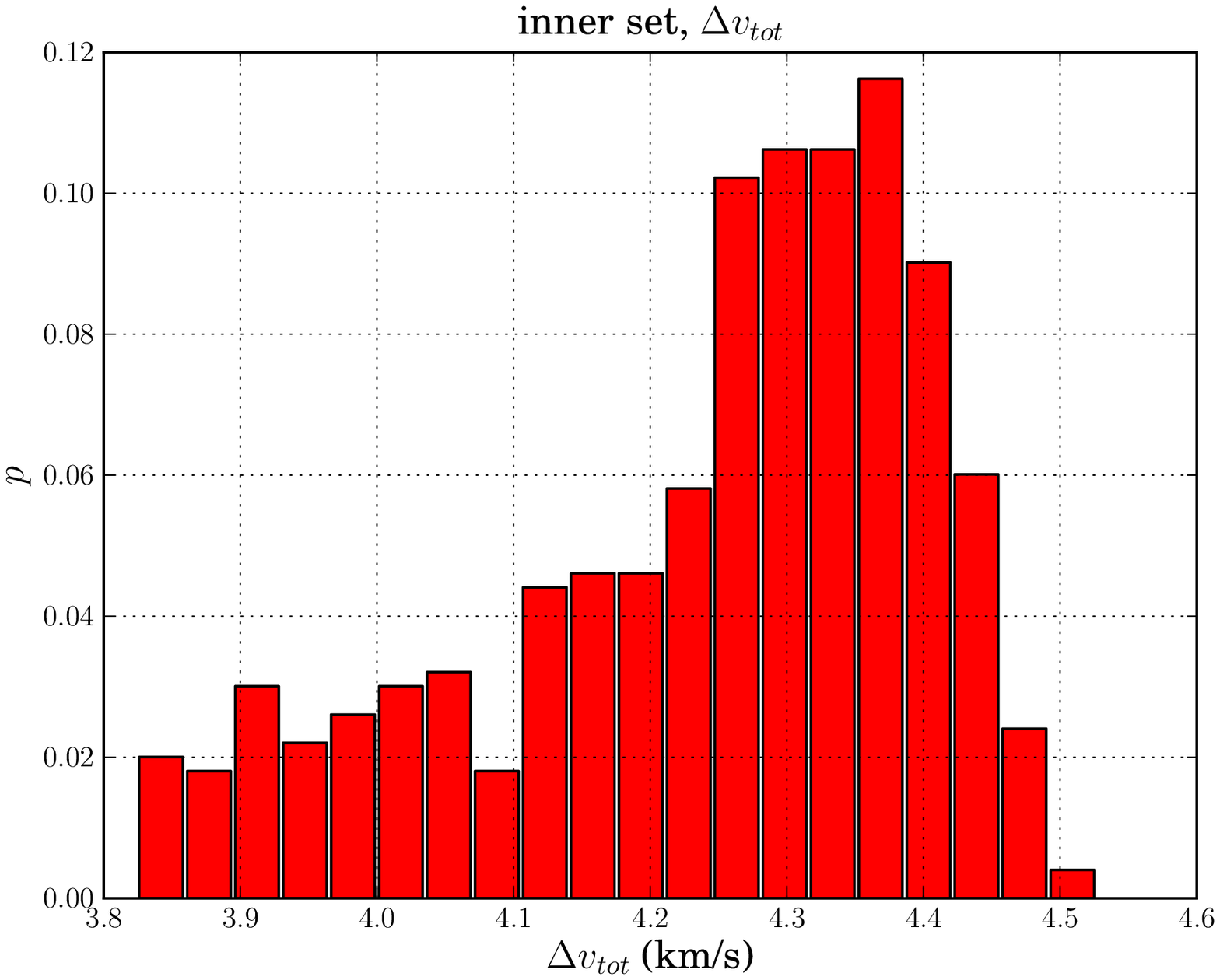}
            }
            \subfigure[]{
               \label{subf:lowen-hists-out-dv}
      	      \includegraphics[width=0.45\textwidth]{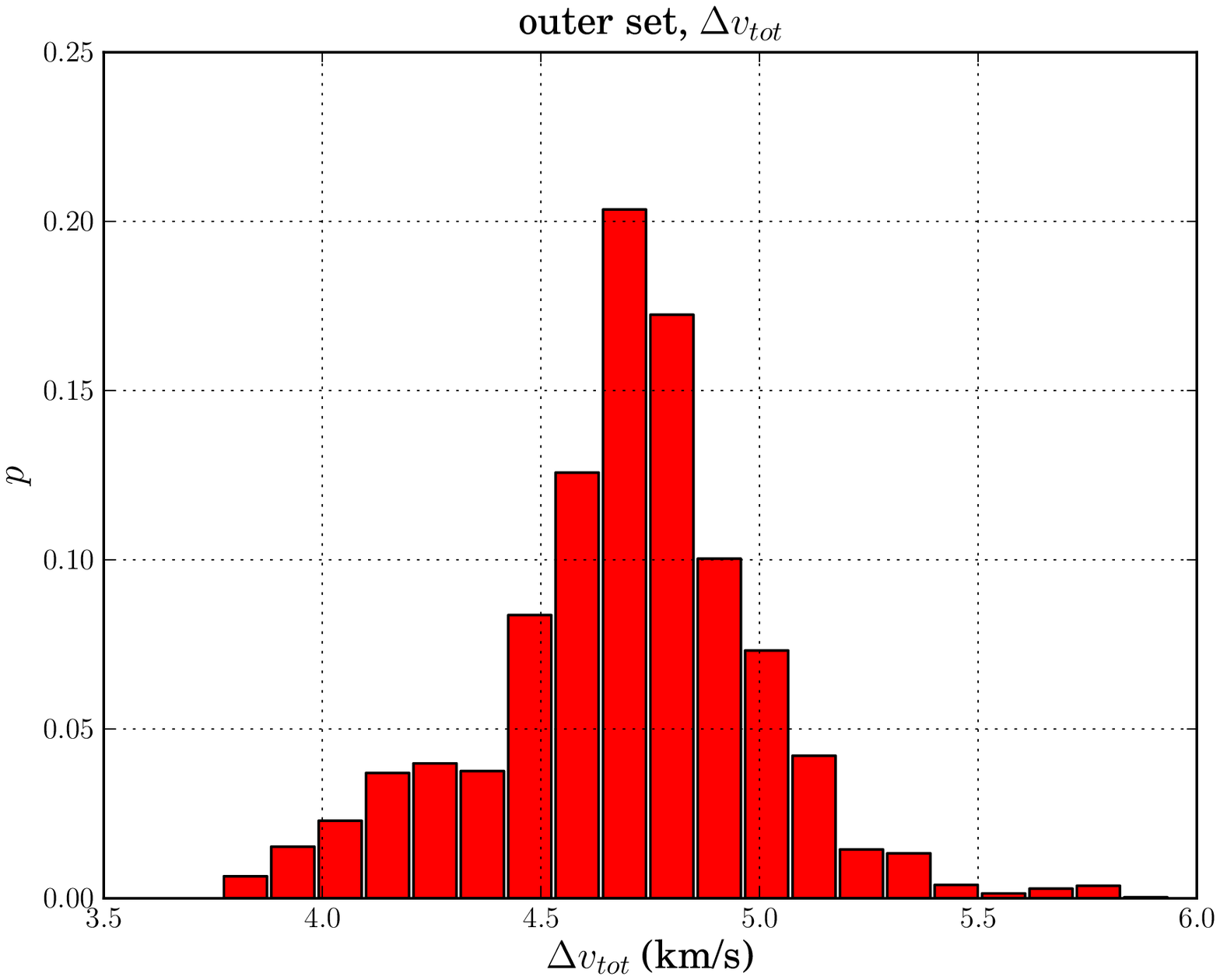}
            }
      	\caption{$\Delta v_{tot}$ of the inner and outer sets
            of transfers. The count is normalized so that the
            sum is unitary.
            \subref{subf:lowen-hists-in-dv} inner set, $499$ orbits,
            $\Delta v_1<3.10\textnormal{ km s}^{-1}$;
            \subref{subf:lowen-hists-out-dv} outer set, $3538$ orbits,
            $\Delta v_1>3.14\textnormal{ km s}^{-1}$}
         \label{fig:lowen-hists}
      	\end{figure*}

         The plots of Figure \ref{fig:lowen-hists} illustrate the $\Delta v$
         distribution of two sets of orbits of interest. The ``inner set'' is
         composed of all the transfers with $\Delta v_1<3.10\textnormal{ km
         s}^{-1}$ (the value of the mean Hohmann transfer), while the orbits
         of the ``outer set'' have $\Delta v_1>3.140\textnormal{ km s}^{-1}$.
         Note that these two sets contain both low-energy and high-energy orbits.

         For both sets the majority of the orbits have high $\Delta v_{tot}$, greater
         than $4.0\textnormal{ km s}^{-1}$, indicating high-energy orbits. However
         the left tails of both distributions extend down to
         $3.8\textnormal{ km s}^{-1}$, where the low-energy orbits are.

\begin{figure*}%[!hp]
            \centering
            \subfigure[]{
               \label{subf:normal}
            \includegraphics[width=0.7\textwidth]{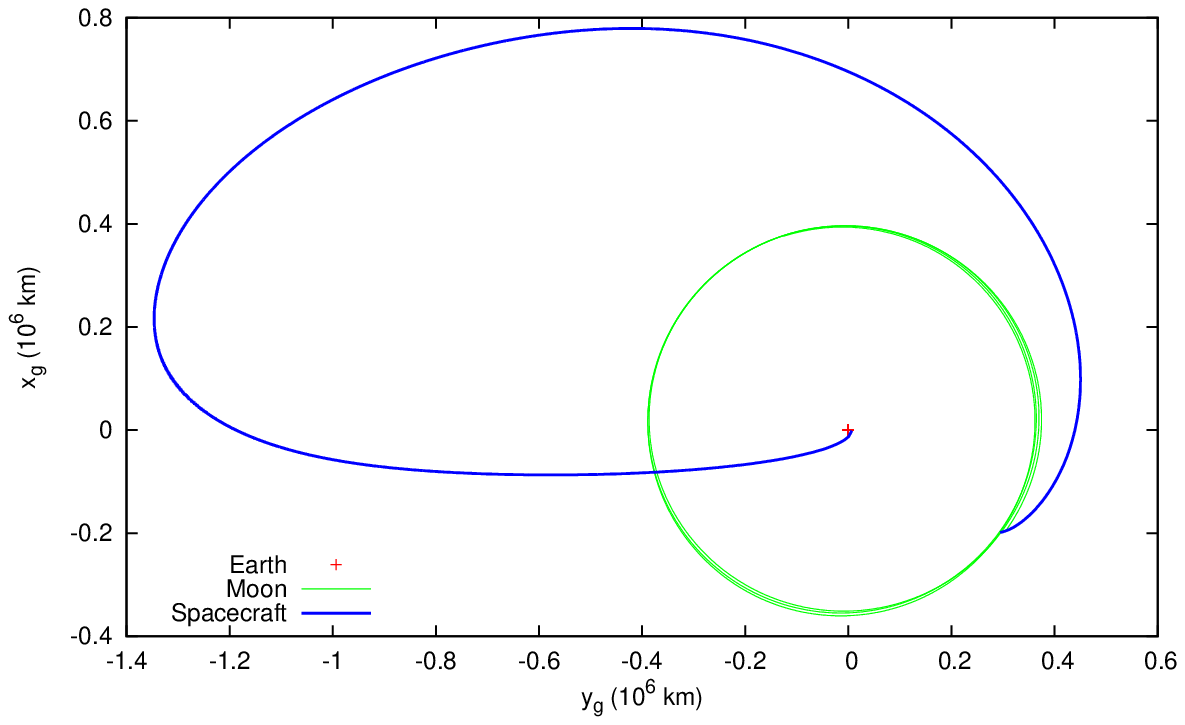}
            }
            \subfigure[]{
               \label{subf:nu-30}
      	      \includegraphics[width=0.7\textwidth]{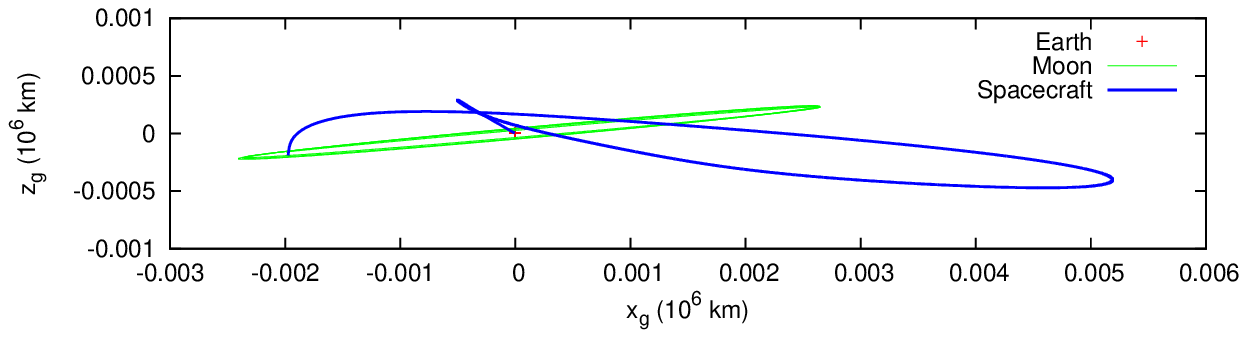}
            }
            \subfigure[]{
               \label{subf:nu-90}
      	      \includegraphics[width=0.75\textwidth]{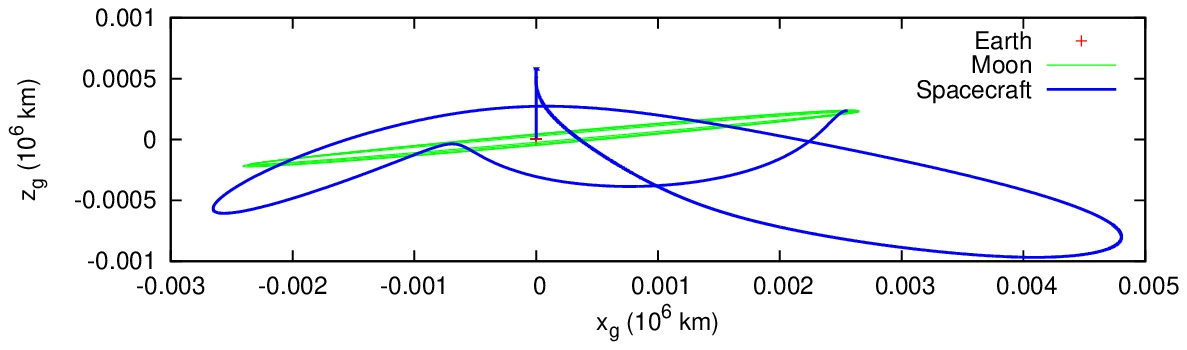}
            }
            \subfigure[]{
               \label{subf:nu90}
      	      \includegraphics[width=0.75\textwidth]{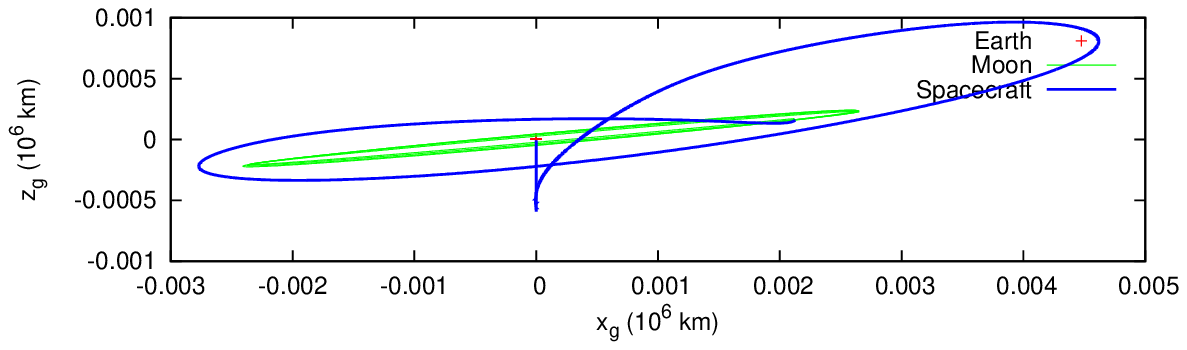}
            }
            \subfigure[]{
               \label{subf:old_innere}
         	   \includegraphics[width=0.45\textwidth]{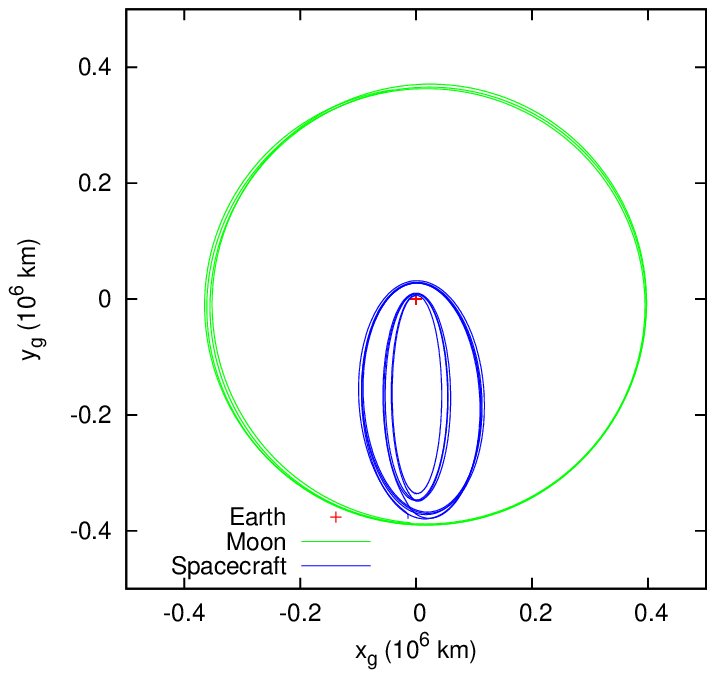}
            }
            \subfigure[]{
               \label{subf:old_innerm}
      	      \includegraphics[width=0.45\textwidth]{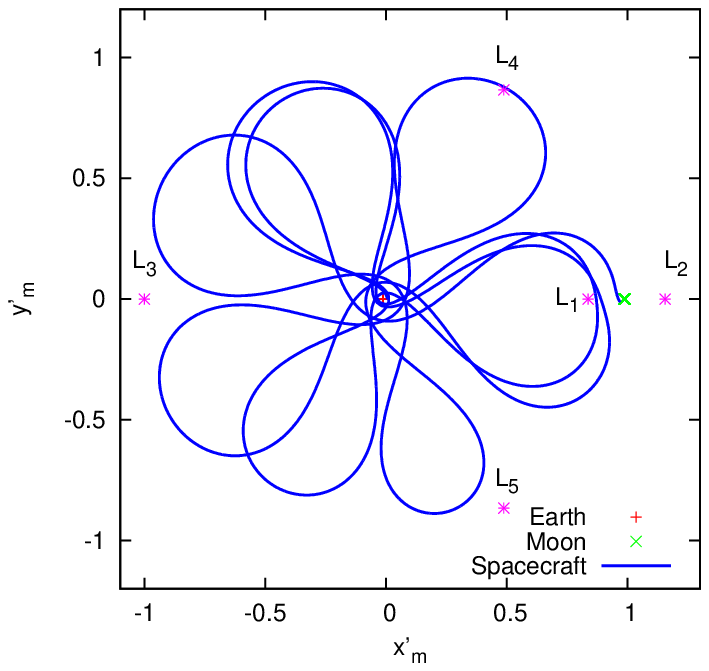}
            }
      	\caption{\subref{subf:normal}: Typical outer transfer, $i=0$, projection on the
            $x$-$y$ plane; \subref{subf:nu-30}: $x$-$z$ plane projection of another outer
            transfer, $i=30$; \subref{subf:nu-90}: outer transfer, $i=90$, $\Omega=0^{\circ}$,
            $x$-$z$ plane; \subref{subf:nu90}: outer transfer, $i=90$, $\Omega=180^{\circ}$,
            $x$-$z$ plane; \subref{subf:old_innere}, \subref{subf:old_innerm}: inner transfer
            in the inertial and fixed-Earth-Moon frames respectively, $i=28.55$, both in
            their $x$-$y$ planes.}
      	\label{fig:examples}
      	\end{figure*}

         \paragraph{Examples of low-energy orbits.}
         Five different low-energy transfers found in the simulation are shown in Figure
         \ref{fig:examples}. Orbits \ref{fig:examples}\subref{subf:normal} to \subref{subf:nu90}
         are outer transfers. \ref{fig:examples}\subref{subf:normal} has $i=0$ from
         the start, while the following three (shown projected on a plane normal to
         the ecliptic) have different initial inclinations. Note that when displayed from
         above the ecliptic, these four transfers all have essentially the same
         shape. Their independence to the initial inclination is due to the nature
         of the sun's tidal force, described in Section \ref{theory}, constantly
         pulling towards the ecliptic plane.

         Figures \ref{fig:examples}\subref{subf:old_innere} and
         \ref{fig:examples}\subref{subf:old_innerm} show the same inner transfer
         in the inertial and in the Earth-Moon rotating frame. The sudden changes
         in the semi-major axis and eccentricity, due to resonance hops, are
         clearly visible.

         \begin{figure*}
            \centering
      	   \includegraphics[width=0.9\textwidth]{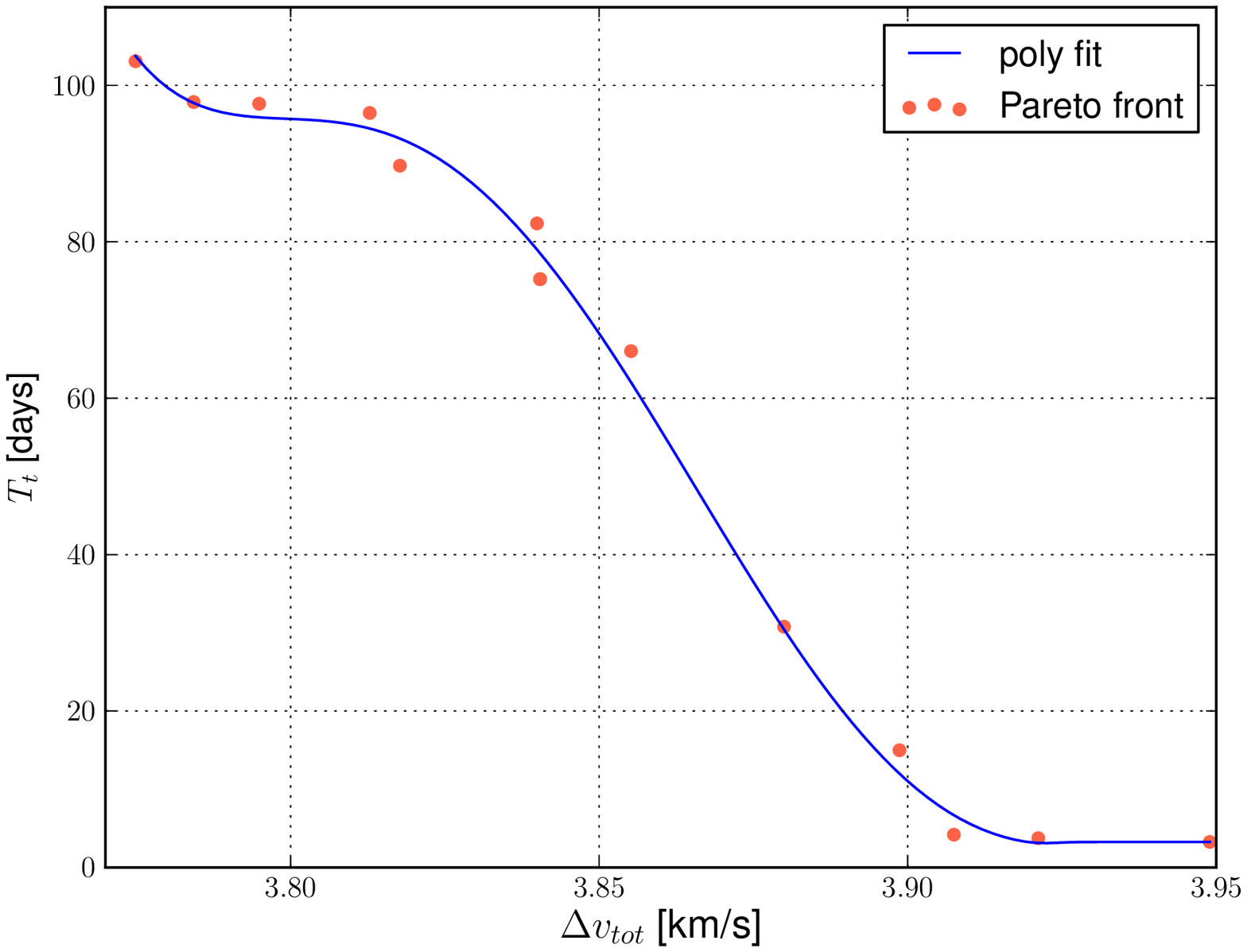}
            \caption{The Pareto front in the $\Delta v_{tot}$-$T_t$ plane and its
                     5$^{\textrm{th}}$ order spline interpolation.}
      	   \label{fig:pareto_fit}
         \end{figure*}

	%%%%%%%%%%%%%%%%%%%% SECTION
   \section{Application of Multi-Objective Optimization Methods}
      \label{opti}
      Like most modern day design problems, the choice of an optimal trajectory
      for a space mission has to take into account a number of contrasting
      objectives, and thus belongs to the category called \emph{multi-objective
      optimization} (MOO).

      MOO studies the functional relation between the \emph{search space},
      that is the space of all the problem parameters, and the
      \emph{objective space}, composed by the functions that quantify all
      the objectives. Every objective is translated into one
      ``objective function'' \citep{miettinen2001}.

      If there is only one objective, the best solution is the one maximizing
      or minimizing a suitable objective function, depending on the nature of
      the problem. When the number of objectives is greater than one, however,
      the conflict between them makes some kind of trade-off necessary.
      The concept of \emph{domination} of one solution over another
      is introduced to identify which is preferable (or at most equal)
      in terms of all the objectives at the same time.
      The set of all solutions that are not dominated by any other solution,
      and therefore best under at least one objective, is called Pareto frontier
      or Pareto-optimal set \citep{miettinen2001}.

      \subsection{MOO for Earth-Moon transfers}
         Normally, multi-objective optimization programming starts from the search
         space and uses an algorithm to find the optimal points in the
         objective space. Such operation is \emph{not} performed in this
         work because a different approach has been chosen for the reasons detailed in
         the preceding sections. Nevertheless, the production of a large number of
         transfer orbits, like the one performed in this work, allows for the population
         of the objective space, where the
         objective functions are the $\Delta v_{tot}$ and the transfer duration $T_t$.
         The Pareto set for this problem can be drawn with sufficient accuracy, as shown
         in Figure \ref{fig:pareto_fit}.

         After the identification of the Pareto-optimal transfers, the next step in
         planning the mission is the investigation of the possible best choice of a
         single solution. This
         task is not easily automated because it depends, obviously, on the
         specific objectives and the restrictions of the mission. Examples of factors
         to consider would be i) the date of launch, ii) whether the mission is manned or
         unmanned, iii) the positions of Sun and Moon at the time of the first impulse, iv)
         the launch base, v) the launcher type and the method of propulsion, etc. All of these
         may be different for each mission project.

         However, the choice of an optimal transfer can be simplified by means of
         multi-objective optimization techniques. In the next subsections this is done
         for the $2$-objective problem of minimizing both the transfer duration and the total
         $\Delta v$, using part of the $7$-dimensional search space formed by the
         initial spatial and velocity coordinates and the date of the first impulse.

         \begin{figure*}
            \centering
      	   \includegraphics[width=\textwidth]{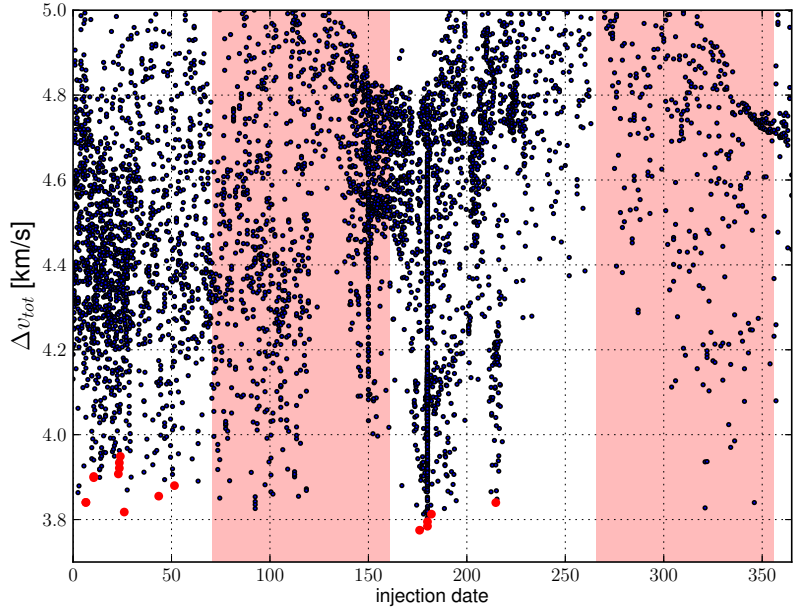}
            \caption{Position of the points of the Pareto front (red dots) in the whole distribution
            by date of the orbits. The ``unfavorable'' quadrants
            ($1^{\textnormal{st}}$ and $3^{\textnormal{rd}}$) are shown on a
            darker background.}
      	   \label{fig:pareto_date}
         \end{figure*}

         \begin{figure*}
            \centering
      	   \includegraphics[width=\textwidth]{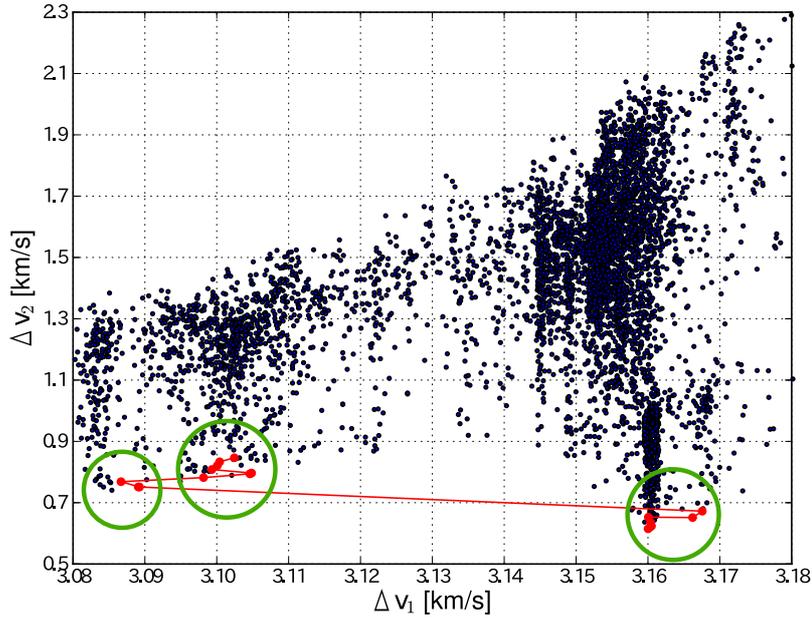}
            \caption{The Pareto front (dots joined by the red line) in the $\Delta v_1$-$\Delta v_2$ plane.
            The line between the points shows the order in which they appear on
            the front (corresponding to increasing values of $\Delta v_{tot}$).
            The points occupy three distinct areas in the plane, highlighted by
            circles, corresponding to inner, Hohmann and outer transfers
            respectively from left to right.}
      	   \label{fig:pareto_dv12}
         \end{figure*}

         \begin{figure*}
            \centering
      	   \includegraphics[width=\textwidth]{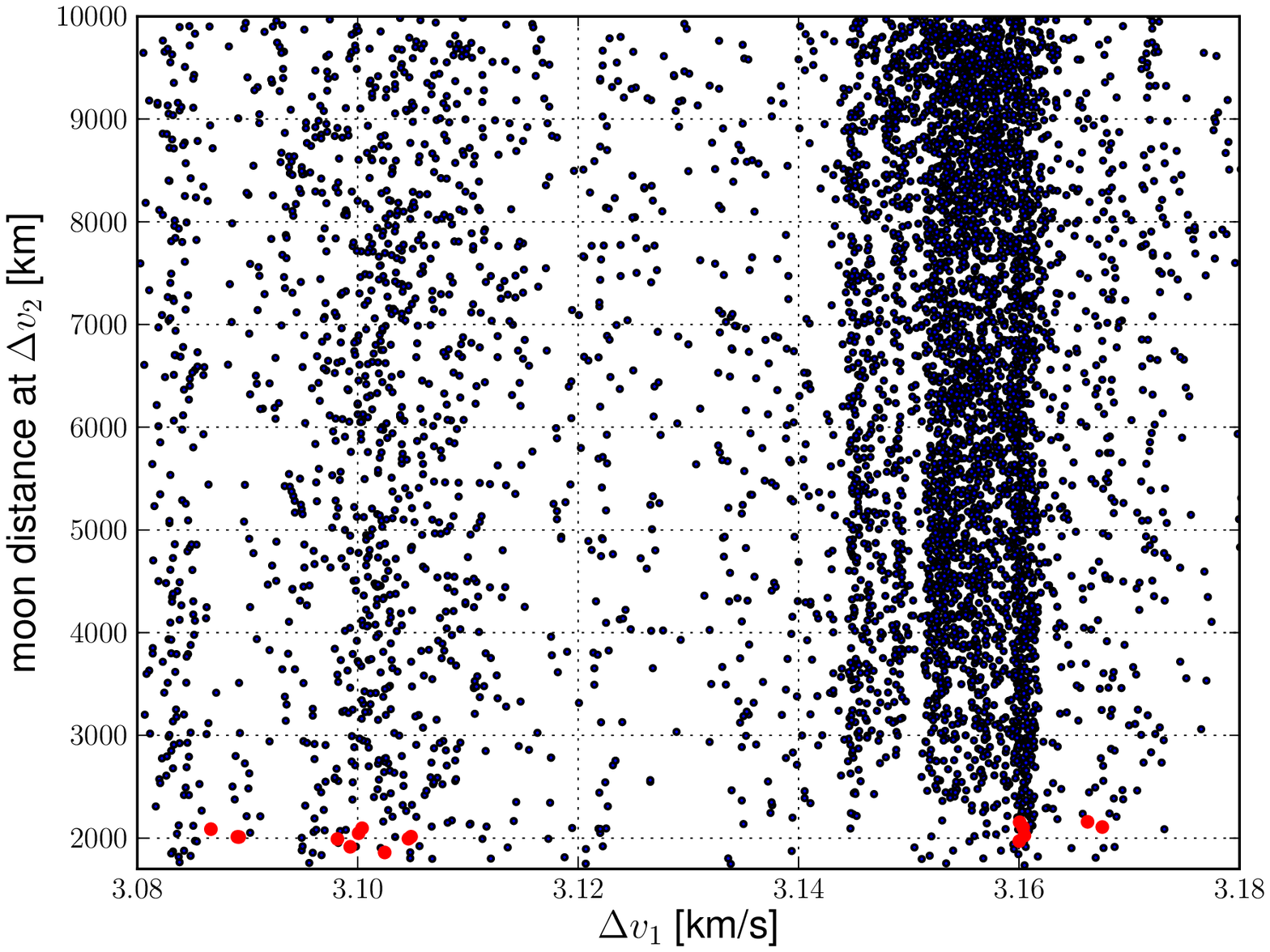}
            \caption{The Pareto front (red points) in the distribution of distances
            from the Moon at injection, in relation to $\Delta v_1$.}
      	   \label{fig:pareto_mdist}
         \end{figure*}

      \subsection{Mapping of the Pareto Front Back to the Search Space}
         As a first step we analyse the location of the Pareto-optimal
         transfers in the space of the initial parameters.
         
         If the Pareto front is a continuous curve, it is expected that also its
         image in the search space is a continuous curve.
         Obviously, the more populated the Pareto front is, the more clearly
         the mapped set will be recognizable as a line.  The front shown in
         Figure \ref{fig:pareto_fit} is determined with $13$ points only, and it is obtained
         from a partial exploration of the search space. In these conditions it
         is anyway possible to deduce valuable, although basic, information.

         Figure \ref{fig:pareto_date} shows the position of the optimal points (in red) in the $\Delta
         v_{tot}$-``starting date'' plane. The periods of time in which $\alpha$ is in its
         four quadrants is shown in white ($2^{nd}$ and $4^{th}$) and colored
         ($1^{st}$ and $3^{rd}$) background.  All the optimal orbits
         fall inside the ``favorable'' ($2^{nd}$ and $4^{th}$) quadrants. 

         Figure \ref{fig:pareto_dv12} shows the same plot as Figure
         \ref{fig:dv12}, that is, the relation between
         $\Delta v_1$ and $\Delta v_2$, but with the points corresponding to the
         Pareto front highlighted in red. As detailed in the previous
         sections, the $\Delta v_1$ is directly related to the apogee distance.
         In the plot, the points corresponding to the Pareto set are grouped
         into three distinct areas equivalent to the inner transfers ($\Delta
         v_1<3.1\textnormal{ km s}^{-1}$ i.e.  apogees inside the orbit of the Moon), the Hohmann
         transfers ($\Delta v_1=3.1\textnormal{ km s}^{-1}$, i.e. apogees equal to the Earth-Moon
         distance) and the outer transfers ($\Delta v_1>3.14\textnormal{ km s}^{-1}$, apogees longer
         than $7\times10^5$ km).

         The plot in Figure \ref{fig:pareto_mdist} is the distribution, again as a function of the
         $\Delta v_1$, of the distance between the spacecraft and the moon during
         the capture maneuver $\Delta v_2$. The general distribution
         of orbits is random, as explained in the previous sections of this paper, 
         and all the optimal transfers have capture
         distances close to $2000$ km, which is the value that minimizes $\Delta v$ for the 
         final circularization \footnote{In fact the final
         orbit around the Moon has a radius of $2000$ km.}. This is not surprising, because
         that the final circularization maneuver contributes significantly
         to the total $\Delta v$.

      \subsection{Scaling Curve}
\begin{figure*}
            \centering
      	   \includegraphics[width=\textwidth]{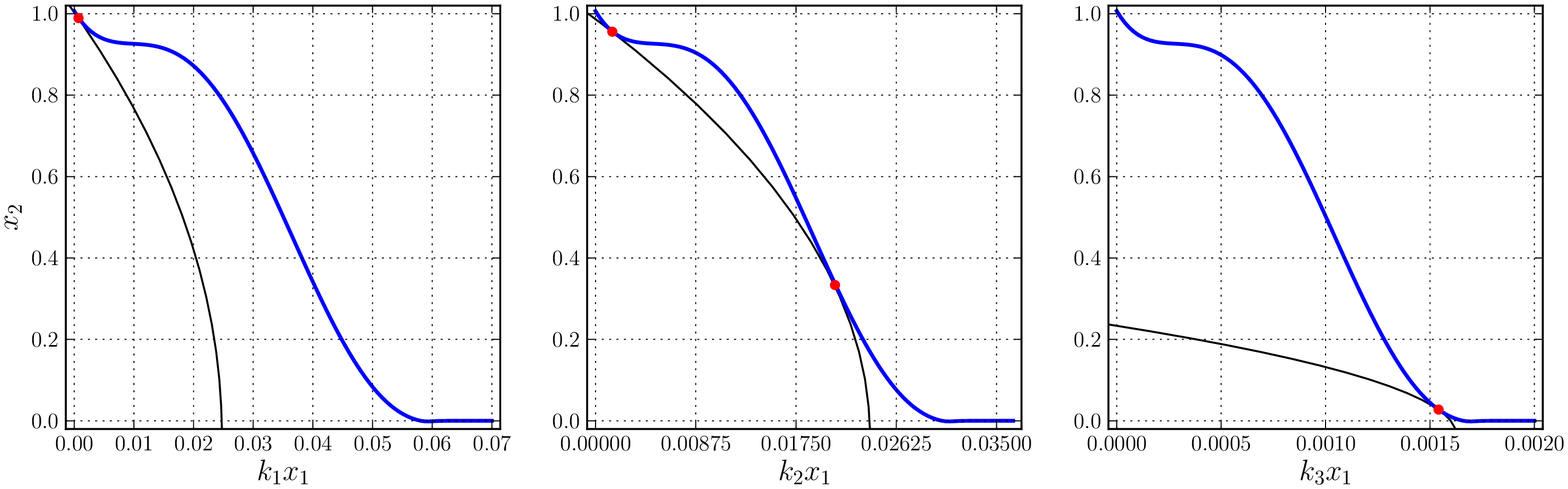}
            \caption{Examples of three different weights for the objective
            functions and the consequent different optimal solutions. The
            axes are normalized for easier handling. Here the $x_1$-axis
            is deformed for easier viewing, so that the front appears to have the same shape,
            while the circumferences of constant distances (in black) appear to be
            ellipses with different eccentricities.}
      	   \label{fig:pareto3scales}
         \end{figure*}

         As was pointed out earlier, there is no algorithm capable of giving
         a single optimal solution for all mission types, and the final
         decision rests with the humans performing the optimization.
         Nevertheless, there are several techniques that aid the
         decision-maker in his choice.
         One of these is the ``weighting method'', in which a weight is given to
         each objective based on its perceived importance. A function
         expressing the distance between the optimal points and a previously
         selected ``ideal'' or ``utopia'' point is minimized. Here we use the
         simple euclidean distance, or $L_{(2)}$ norm, between a generic point
         $\left(x_1(p),x_2(p)\right)$ and the utopia point $\left(x_1^*,x_2^*\right)$. The $L_{(2)}$
         norm is defined as:
         \begin{equation}
            d_{12}=\left[\left(x_1(p)-x_1^*\right)^2+\left(x_2(p)-x_2^*\right)^2\right]^{\frac12}
         \end{equation}
         The utopia point adopted here is the origin, ($\Delta v_{tot}=0$,
         $T_t=0$).  Figure \ref{fig:pareto3scales} shows how different weighting
         choices deform the front and result in different optimal points being closest to the 
         utopia point (in normalized units).

         Clearly, the weighting method reduces the problem to a single-objective
         minimization process. However, it has the drawback of having to choose 
         the weights before seeing the respective solution. In some cases a small 
         change in the weights can result in large differences in the solution, and the decision-maker
         has a hard time finding a clear answer. This is partly solved by the
         ``scaling method'' developed by \citet{kasprzak2001}. This method takes each point of
         the Pareto front and computes the weighting (or scaling) that would be
         necessary to make that solution the closest one to the utopia point.

\begin{figure*}
            \centering
      	   \includegraphics[width=\textwidth]{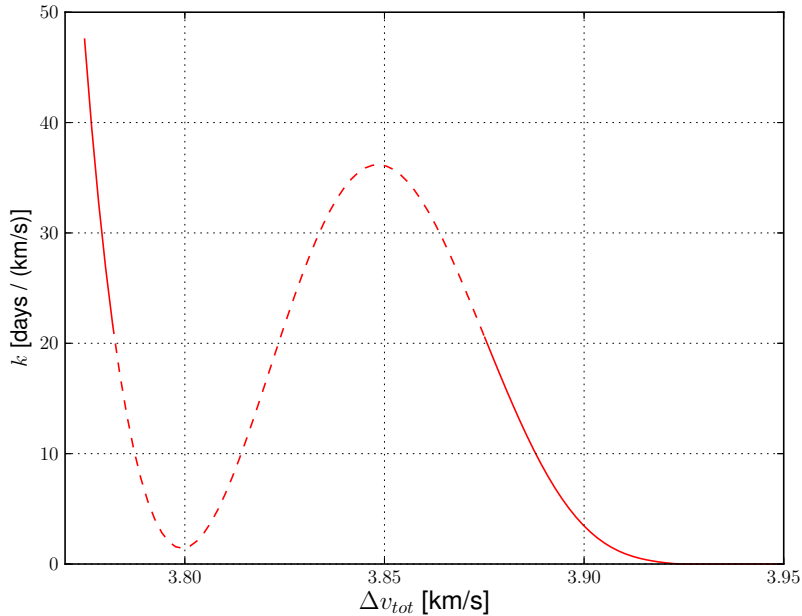}
            \caption{The scaling curve resulting from the Pareto front in Figure
                     \ref{fig:pareto_fit} and a utopia point at the origin. The
                     dashed part denotes non-optimal solutions.}
                     \label{fig:scaling_curve}
         \end{figure*}

         The general procedure used in the scaling method is as follows:
         \begin{itemize}
            \item fit the Pareto optimal set with a high-order polynomial or a
               spline curve;
            \item for each point in the curve, find the local slope $m$;
            \item compute the rescaling necessary to make the slope normal to
               the distance vector that must be minimized; the relation is
               $k=-m^{-1}$.
         \end{itemize}
         The result is a curve relating one of the objective functions to the scaling
         that leads to optimality. For example, by looking at Figure \ref{fig:pareto3scales},
         we see that the
         curve would have (in the normalized units) a value of $k_1=0.07$ for
         the red point shown in the first graph, $k_2=0.036$ for the two points
         in the second graph and $k_3=0.002$ for the point in the third graph.
         The full curve, expressed in the original, physical units, is shown in Figure
         \ref{fig:scaling_curve}.

         The scaling curve obtained in this way facilitates the decision-maker's
         job, because he now has information covering the whole spectrum of solutions,
         including the sensitivity to a change of weighting. The fact that in some
         cases more than one Pareto point corresponds to the same $k$ scaling means
         that there are multiple solutions with the same distance to the utopia point
         and thus identically optimal, like for $k_2$ in Figure \ref{fig:pareto3scales}.
         However, the concave part of the front is never optimal in the weighting method.
         Its points never minimize the distance to the utopia point. This can be seen in
         the middle plot in Figure \ref{fig:pareto3scales}, where the whole arc between the
         two red dots is clearly unreachable by the circumference. All points of this kind
         are shown as a dashed curve in
         Figure \ref{fig:scaling_curve} and should be excluded from the optimization
         procedure.
         
         \paragraph{Practical application.}
         The scaling $k$ in Figure \ref{fig:scaling_curve} can be interpreted
         physically as the number of days of transfer that have, for the
         specific purposes of the mission at hand, the same impact as a $\Delta
         v_{tot}$ of $1\textnormal{ km}/\textnormal{s}$.

         For example, on a manned mission, 1 day of flight might be as expensive, in terms of
         security and costs, as $1\textnormal{ km}/\textnormal{s}$ of the total
         $\Delta v$. Using Figure \ref{fig:scaling_curve}, we look for the point
         in the curve that has $k=1\textnormal{ days}/(\textnormal{km }s^{-1})$.
         Since we have the whole scaling curve, our choice of $k$ need not be
         only one, but can be a range of values. In this case we see that the
         best solution would be for $\Delta v_{tot}>3.91\textnormal{ km}/\textnormal{s}$.
         From this, using the Pareto front, we find that the best transfer orbits are
         the traditional Hohmann transfers.

         On the other hand, transfer duration may have
         a much lower cost for an unmanned mission with science purposes, with
         a desired scaling $k\simeq25\textnormal{ days} / (\textnormal{km }s^{-1})$.
         In this case the optimal choice would be a low-energy orbit with
         $\Delta v_{tot}\simeq3.78\textnormal{ km}/\textnormal{s}$ and (from the
         Pareto front) $T_t\simeq95$ days.

         After locating an optimal transfer (or several candidates) on the front,
         mapping back to the search space gives its initial parameters like launch
         date, inclination and $\Delta v_1$.

	%%%%%%%%%%%%%%%%%%%% SECTION
   \section{Conclusions}
   \label{concl}
      In this paper we study the existence and characteristics of low- and high energy transfer
      orbits from the Earth to the Moon in the combined field of Earth, Moon and Sun.
      A wide-ranging and numerically precise study of these orbits contributes to a
      better understanding of the orbital dynamics. Among these transfers, low-energy
      orbits give clear advantages in terms of mission costs. A result
      of this work is that it is easy to obtain successful low-energy orbits just by
      properly setting orbital initial conditions, without resorting to mid-course
      maneuvers as, for instance, was the case for the Hiten spacecraft
      \citep{belbruno-hiten}.
      
      On the other side, this work has not the scope of a deep understanding of the
      underlying dynamical characteristics of these orbits, which is better studied
      by semi-analytical or geometrical approaches, like for instance the elegant
      techniques involving invariant manifolds
      \citep{kolomaro2000,gomez-lo2002,kolomaro-book}. In particular, it was shown how
      it is possible using invariant manifold techniques to construct low energy
      transfer paths to the Moon \citep{kolomaro2001}.
      
      The existence of such orbits relies strongly on the help of the Sun, and the
      statistics show that a specific range of positions is required for the Sun
      for the practical realization of low-energy orbits, confirming previous results
      \citep{teofilatto2001}.  Another relevant result of this paper
      is the information about the range of apogee distances which most likely
      produce two-impulse outer low-energy transfers, which is found to be
      $1.23\times10^6\textnormal{ km} \leq r_a \leq 1.31\times10^6\textnormal{ km}$.
      Moreover, the simulations show that the outer orbits are capable of reaching
      the Moon independently of the inclination of the initial parking orbits, unless
      they are retrograde. 
      
      As a relevant final point, we discussed the use of multi-objective optimization
      (MOO) methods for the choice of appropriate transfer orbits.  The acquired data
      allow us to draw the shape of the Pareto frontier, which is the fundamental
      element of a complete MOO analysis.  After determining the Pareto
      front (set of all the optimal points in the objective space) it is possible
      to map back those points onto the search space (that of initial
      conditions) to get the optimal orbit initial conditions (launch date,
      inclination and first velocity impulse $\Delta v_1$).
      
      While, clearly, a
      general technique of resolution of the problem of finding a single optimal
      transfer orbit is not viable because of the variety of contrasting
      objectives, whose specific relevance is highly mission-dependent, we sketched
      that a suitable first approach is that of reducing the MOO to a two-objective
      problem. The two objectives are the total impulse $\Delta v_{tot}$ and the
      transfer time duration $T_t$. Also in this simplified frame a high degree of
      indeterminacy remains due to that the scope of the mission suggest whether
      higher relevance is to be given to the amount of injected impulse or to the
      transfer time.
      
      The final tool used in the selection of a single optimal transfer is the scaling
      curve. The scaling curve shows the best choice given any $T_t$--$\Delta v_{tot}$
      preference factor $k$, which is mission dependent. We explore two different hypothetical
      cases of manned and unmanned missions. We find, choosing a $k=1\textnormal{
      day}/(\textnormal{ km s}^{-1})$ for the sake of a manned mission,
      that optimal orbits are those with $\Delta v
      >3.91\textnormal{ km s}^{-1}$ which
      coincide with usual Hohmann orbits, while the request of $k \sim 25\textnormal{
      days}/(\textnormal{ km s}^{-1})$, quite reasonable for unmanned missions, leads
      to a low energy orbit as optimal, with $\Delta v \simeq3.78\textnormal{
      km}/\textnormal{s}$ and $T_t
      \simeq95$ days.

%%%%%%%%%%%%%%%%%%%%%%%%%%%%%%%%%%%%%%%%%%%%%%%%%%%%%%%%%%%%%%

\begin{acknowledgements}
	We thank prof. Teofilatto (Sapienza, University of Roma, Italy)
   for useful discussions during the preparation of this paper.
	\end{acknowledgements}

	%%%%%%%%%%%%%%%%%%%% REFERENCES
	\bibliographystyle{spbasic}      % basic style, author-year citations
   \bibliography{thebib}
\end{document}